# Large energy soliton erbium-doped fiber laser with a graphene-polymer composite mode locker

Han Zhang[1], Qiaoliang Bao[2], Dingyuan Tang[1]*, Luming Zhao[1], and Kianping Loh[2]

1. School of Electrical and Electronic Engineering, Nanyang Technological University, Singapore 639798

2. Department of Chemistry, National University of Singapore, Singapore 117543

* Corresponding author: edytang@ntu.edu.sg
Website: http://www3.ntu.edu.sg/home2006/ZHAN0174/

Due to its unique electronic property and the Pauli Blocking Principle, atomic layer graphene possesses wavelength-independent ultrafast saturable absorption, which can be exploited for the ultrafast photonics application. Through chemical functionalization, a graphene-polymer nanocomposite membrane was fabricated and firstly used to mode lock a fiber laser. Stable mode locked solitons with 3 nJ pulse energy, 700 fs pulse width at the 1590 nm wavelength have been directly generated from the laser. We show that graphene-polymer nanocomposites could be an attractive saturable absorber for high power fiber laser mode locking.

Recently, passive mode-locking of fiber lasers using single-wall carbon nanotubes (SWCNTs) as saturable absorbers has attracted considerable attention. S. Y. Set et al firstly reported fiber laser mode locking using SWCNTs [1]. It was shown that SWCNTs have broadband saturable absorption and a fast absorption recovery time, which could be used as a saturable absorber for laser mode locking. Conventionally, passive mode locking of a laser is achieved with a semiconductor saturable absorption mirror (SESAM). However, fabrication of SESAMs requires expensive and complex epitaxial growth technique. In contrast, saturable absorbers based on carbon nanotubes could be easily fabricated [1-4]. Following S. Y. Set et al, mode locking of fiber lasers using SWCNTs or its polymer composites has been intensively investigated. Worth of mentioning are the works by Y. W. Song et al., who have achieved in an erbium-doped fiber laser picosecond pulse emission with a record high pulse energy of ~6.5 nJ [5], and F. Wang et al, who have demonstrated wideband tuning of a mode locked erbium-doped fiber laser [6]. In addition, K. Kieu and F. W. Wise have shown mode locking of an all-normal dispersion Yb fiber laser using a SWCNTs-based absorber, and generated pulses with 1.5 ps pulse duration and 3 nJ pulse energy [7]. Recently, mode locking of solid-state lasers using SWCNTs has also been demonstrated [8].

A drawback of SWCNTs-based saturable absorbers is that SWCNTs tend to form bundled and entangled morphology, which causes strong scattering and thus strong non-saturable losses. In addition, under large energy ultrashort pulse radiation multi-photon effect induced oxidation occurs [9], which degrades the long-term stability of the absorber. In this letter we report on the soliton mode locking of an erbium-doped fiber laser using a graphene-polymer nanocomposite membrane as the mode locker. Graphene is a single two-dimensional (2D) atomic layer of carbon atom arranged in a hexagonal lattice. Although as an isolated film it is a zero bandgap semiconductor, we found that like the SWCNTs, graphene also possesses saturable absorption [10]. In particular, as it has no bandgap, its saturable absorption feature is wavelength independent. It is potentially possible to use graphene or graphene-polymer composites to make a wideband saturable absorber for laser mode locking. Furthermore, comparing with the SWCNTs, as graphene has a 2D structure it should have less surface tension, therefore, much higher damage threshold. Indeed, in our experiments with an erbium-doped fiber laser we have first demonstrated self-started mode locking of the laser with a grapheme-polymer membrane as the saturable absorber, and achieved stable soliton pulses with 700fs pulse width and 3nJ pulse energy.

Our fiber laser is schematically shown in Fig. 1a. A piece of 5.0 m, 2880 ppm Erbium-doped fiber (EDF) with group velocity dispersion (GVD) of -32 (ps/nm)/km was used as the gain medium, and 23.5 m single mode fiber (SMF) with GVD 18 (ps/nm)/km was employed in the cavity to compensate the normal dispersion of the EDF and ensure that the cavity has net anomalous GVD. The net cavity dispersion is estimated -0.3419 $ps^2$. A 30% fiber coupler was used to output the signal, and the laser was pumped by a high power Fiber Raman Laser source (KPS-BT2-RFL-1480-60-FA) of wavelength 1480 nm. The maximum pump power can be as high as 5 W. A polarization independent isolator was spliced in the cavity to force the unidirectional operation of the ring cavity, and an intra-cavity polarization controller (PC) was used to fine-tune the linear cavity birefringence. An optical spectrum analyzer (Ando AQ-6315B) and a 350 MHz oscilloscope (Agilent 54641A) combined with a 2 GHz photo-detector was used to simultaneously monitor the spectra and pulse train, respectively. A graphene-polymer nanocomposite membrane with a thickness of ~ 10 µm inserted between two ferules was used as the saturable absorber for the laser mode locking. To make the graphene-polymer nanocomposite, graphene nanosheets were produced by chemically reducing the oxidized graphene exfoliated from graphite flakes [11]. The graphene nanosheets were then non-covalently functionalized with 1-pyrenebutanoic acid, succinimidyl ester (PBASE) to improve their solubility in organic solvents (i.e., ethanol and acetone) and compatibility with polymers (i.e., PVDF). The as-produced graphene (2 mg) was further dispersed in ethanol (3mL) and mixed with PVDF solution comprising 1.5 g PVDF dissolved in 10 mL dimethylacetamide/acetone (2:3). The G-PVDF solution was then stirred for 24 hours at 60 ºC in sealed bottle to form wet paste for electrospinning. The electrospinning was carried out in a MECC NANON ELECTROSPINNING SETUP at a bias voltage of 30 kV and feeding rate of 0.5 mL/hour. Fig. 1b is a scanning electron microscopy (SEM) image of the membrane. It shows that the membrane mainly comprises networks of the graphene-filled polymer nanofibers. The inset of Fig. 1b is the photo of a free-standing membrane. Fig. 1c shows the transmission electron microscopy (TEM) image of a graphene-PVDF nanofiber. It reveals that the graphene nanosheets are well dispersed in the polymer matrix without obvious aggregation.

We have also measured the linear and nonlinear absorption of the graphene-polymer membrane. The linear absorption spectra of both graphene-based PVDF nanocomposite and pure PVDF were compared in Fig. 2a , which shows that pure PVDF has a relatively low absorbance of ~35% in the L-band of the optical communication windows while graphene-based PVDF nanocomposites has an enhanced absorbance of ~52%. The nonlinear absorption curve of Fig. 2b measured at the wavelength of $\lambda = 1590$ nm shows that the graphene-based PVDF nanocomposite has a normalized modulation depth of ~28.3 % and a saturable fluence of 0.75 MW/cm$^2$, which is about an order of magnitude smaller than that of the SWCNTs based saturable absorber. Moreover, the insertion loss of the graphene-based PVDF nanocomposite was as low as ~1.5dB. Previous studies have also shown that graphene and graphite thin films have both an ultrafast absorption recovery time constant of ~ 200 fs and a slower recovery constant of 2.5 ps to 5 ps [12].

Mode locking of the laser self-started at a pump power of ~400 mW. Fig. 3a shows the typical optical spectra of the mode locked laser emissions. The spectra have a broad spectral bandwidth with obvious Kelly spectral sidebands, characterizing that the mode locked pulses are optical solitons. The central wavelength of the spectra is at 1589.68 nm, which is in the L-band of the optical communication windows. The 3 dB bandwidth of the spectra is about 5.0 nm. Fig. 3b shows the measured autocorrelation trace of the solitons. It has a Sech-profile with a FWHM width of about 1.07 ps, which, divided by the decorrelation factor of 1.54, corresponds to a pulse width of 694 fs. The time-bandwidth product (TBP) of the pulses is 0.412, showing that the solitons are slightly chirped. We also measured the radio-frequency (RF) spectrum of the mode locking state. Fig. 3d shows a measurement made at a span of 10 kHz and a resolution bandwidth of 10 Hz. The fundamental peak located at the cavity repetition rate of 6.95 MHz has a signal-to-noise ratio of 65 dB. The insert of Fig. 3d shows the wideband RF spectrum up to 1 GHz. The absence of spectral modulation in RF spectrum indicates that the laser operates well in the CW mode-locking regime.

Different from the soliton operation of the erbium-doped fiber lasers mode locked with the nonlinear polarization rotation (NPR) technique or a SESAM, no multiple solitons were formed in the cavity immediately after the mode locking. Instead only one soliton was always initially formed. This experimental result shows that the laser has a much lower effective mode locking threshold than those mode locked with SESAMs, which is traced back to the much smaller saturation fluence of the grapheme-polymer nanocomposite than the SESAMs. The single soliton operation could be maintained in the laser as the pump power was gradually increased to 2 W. Further increasing pump power, pulse breaking occurred. Eventually multiple solitons were formed in the laser. Under multiple soliton operations occasionally harmonic mode locking was also observed. We have focused on the single soliton operation of the laser. The energy of the soliton increased with the pump power. A maximum output power of ~13.1 dBm had been obtained under the pump power of ~2 W, which indicates the single soliton energy as high as ~3 nJ. Experimentally we found that the pump power could be increased to as high as 3.2 W and the laser output power could be as large as 17 dBm. Under the pump and laser operation condition the mode locking of the laser could still maintain for hours, which indicates that the graphene-polymer composite could endure at least an optical fluency as high as 21.4 mJ/cm$^2$. After the operation we had also checked the graphene-polymer composite film using the optical microscopy and found that its morphology was kept intact, which verified its strong thermal stability.

In order to investigate the long-term stability of the mode locking, we have recorded the soliton spectra of the laser in a 4-hour interval over 2-days, as shown in Fig.3a. It shows that the soliton parameters, these are the central wavelength, 3 dB spectral bandwidth, Kelly sideband positions and the spectral peak powers, have kept reasonably unchanged. Experimentally, it was also found that the soliton pulse width could be compressed to ~ 524 fs after passing through a 10 m dispersion compensation fiber (DCF) of GVD ~ -4 (ps/nm)/km. The result shows that the output solitons were negatively chirped. We note that the current experimental results were obtained by the first try of the mode locking technique. It is expected that through further careful design of the laser cavity and optimization on the saturable absorption parameters, mode locked pulses with even larger pulse energy and narrower pulse width could be generated.

In conclusion, we have reported an erbium-doped soliton fiber laser with a graphene-polymer nanocomposite membrane as the mode locker. Self-started mode locking of the laser was first experimentally demonstrated, and stable soliton pulses at 1590 nm with 3 nJ pulse energy and 700 fs pulse width were directly generated from the laser. Our experimental results have clearly shown that a graphene-polymer composite membrane has not only the desired saturable absorption for laser mode locking, but also a large damage threshold. It could be a cost-effective saturable absorber for large energy fiber laser mode locking.

K. P. Loh wishes to acknowledge funding support from NRF-CRP Graphene Related Materials and Devices (R-143-000-360-281).

**References:**

1.  Sze Y. Set, H. Yaguchi, Y. Tanaka, and M. Jablonski, J. Light Technology, **22**, 51-56 (2004).

2.  K. Kieu and M. Mansuripur, Opt. Lett. **32**, 2242-2244 (2007).

3.  Z. Sun, A. G. Rozhin, F. Wang, V. Scardaci, W. I. Milne, I. H. White, F. Hennrich, and A. C. Ferrari, Appl. Phys. Lett. **93**, 061114 (2008).

4.  S. Yamashita, Y. Inoue, S. Maruyama, Y. Murakami, H. Yaguchi, M. Jablonski, and S. Y. Set, Opt. Lett. **29**, 1581-1583 (2004).

5.  Y. W. Song, S. Yamashita, and S. Maruyama, Appl. Phys. Lett. **92**, 021115 (2008).

6.  F. Wang, A. Rozhin, V. Scardaci, Z. Sun, F. Hennrich, I. H. White, W. I. Milne, and A. C. Ferrari, Nature Nanotechnology **3**, 738-742 (2008).

7.  K. Kieu and F. W. Wise, Opt. Express, **16**, 11453-114538 (2008).

8.  K. H. Fong, K. Kikuchi, C. S. Goh, S. Y. Set, R. Grange, M. Haiml, A. Schlatter and U. Keller, Opt. Lett., **32**, 38-40 (2007).

9.  T. R. Schibli, K. Minoshima, H. Kataura, E. Itoga, N. Minami, S. Kazaoui, K. Miyashita, M. Tokumoto, Y. Sakakibara, Opt. Express, **13**, 8025-8031 (2005).

10. Q. L. Bao, H. Zhang, Y. Wang, Z. H. Ni, Y. L. Yan, Z. X. Shen, K. P. Loh and D. Y. Tang, Adv. Func. Mat.: DOI: 10.1002/adfm.200901007 (Published Online Aug 25 2009).

11. W. S. Hummers and R. E. Offeman, J. Am. Chem. Soc. **80**, 1339 (1958).

12. R. W. Newson, J. Dean, B. Schmidt and H. M. van Driel, Opt. Express, **17**, 2326-2333 (2009).

**Figure captions:**

Fig.1: (a) Schematic of the soliton fiber laser. WDM: wavelength division multiplexer. EDF: erbium-doped fiber. PC: polarization controller. SMF: single mode fiber. (b) SEM image of the graphene-polymer nanofiber networks. Inset: a photo of the free-standing graphene-polymer composite membrane. (c) Transmission electron microscopy (TEM) image of a graphene-PVDF nanofiber.

Fig.2: (a) UV-VIS-NIR absorption spectra of graphene-based PVDF nanocomposites and pure PVDF. The inset shows the chemical structure of the functionalized graphene. b) Power dependent nonlinear saturable absorption of graphene-based PVDF nanocomposites.

Fig. 3: Soliton operation of the fiber laser. (a) Multiple soliton spectra measured at a 4-hour interval over 2-days. (b) Autocorrelation traces of the solitons. (c) An oscilloscope trace of the laser emission. (d) The fundamental radio-frequency (RF) spectrum of the laser output. Insert: RF spectrum up to 1 GHz

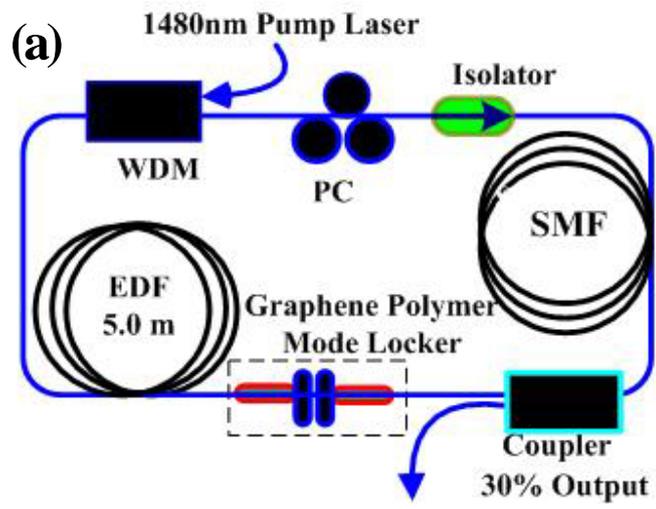

Fig.1 (a): Han Zhang et. al.

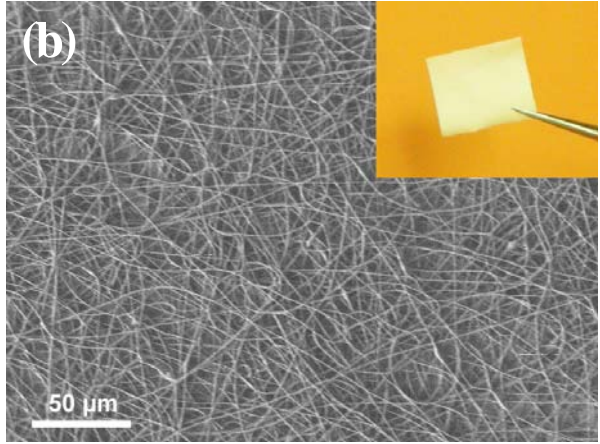

Fig.1 (b): Han Zhang et. al.

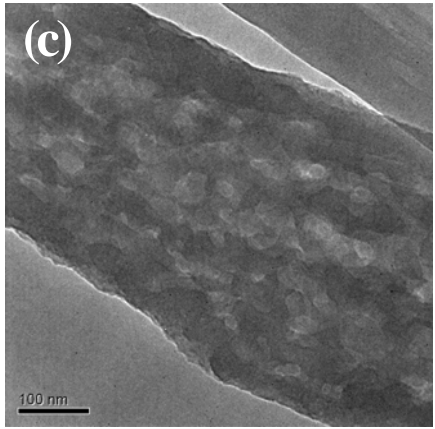

Fig.1(c): Han Zhang et. al.

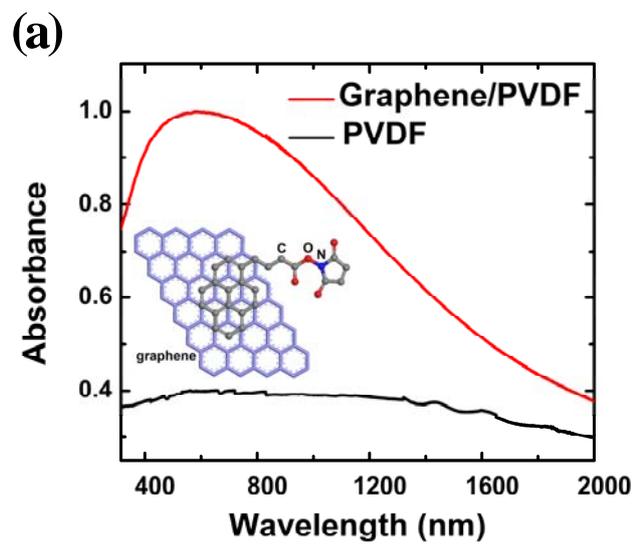

Fig.2 (a): Han Zhang et. al.

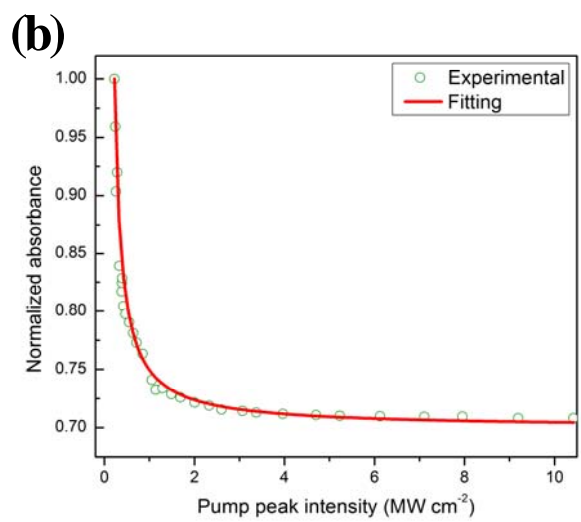

Fig.2 (b): Han Zhang et. al.

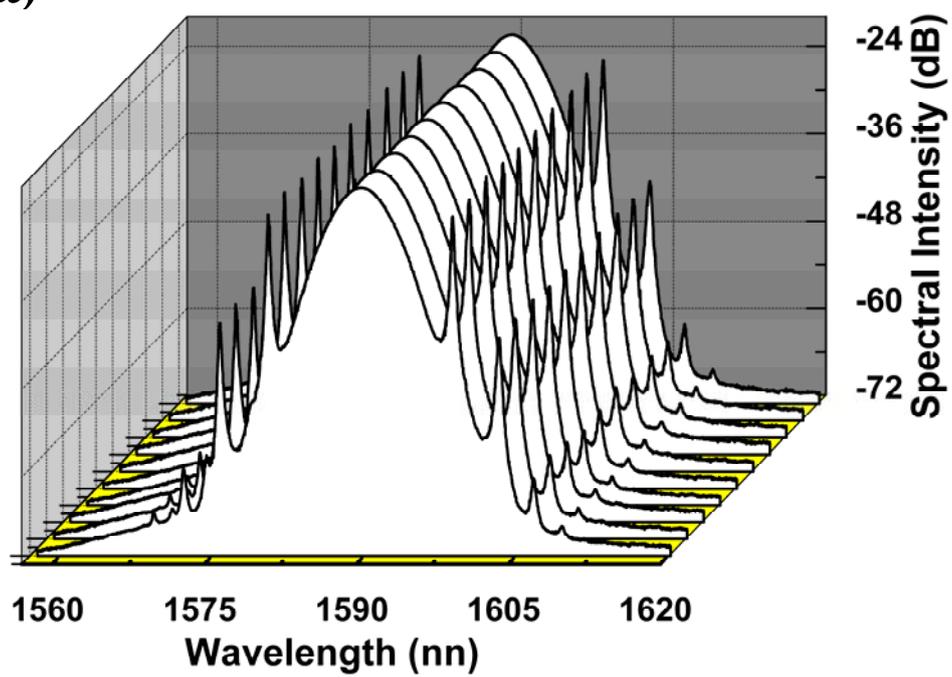

Fig.3 (a): Han Zhang et. al.

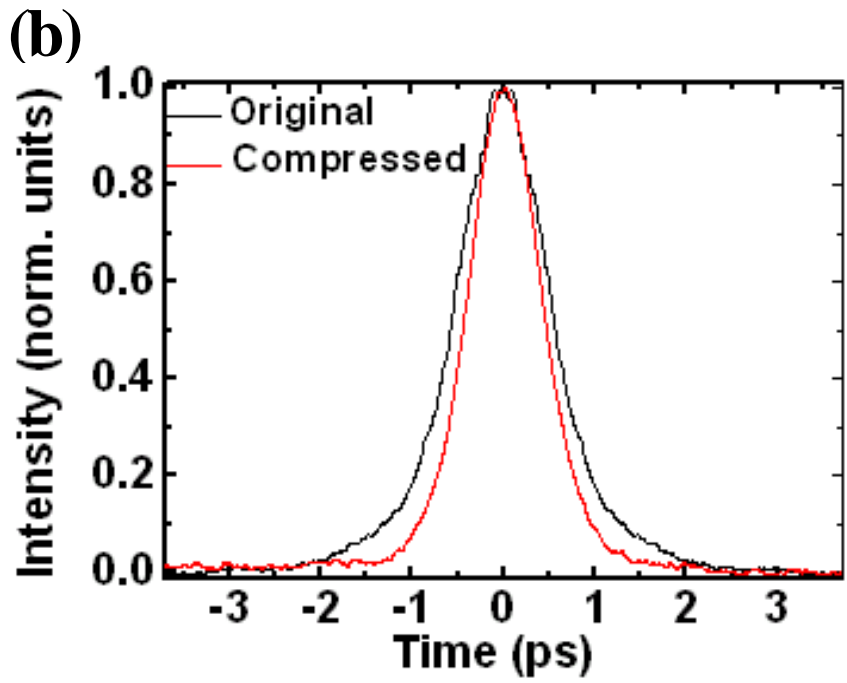

Fig.3 (b): Han Zhang et. al.

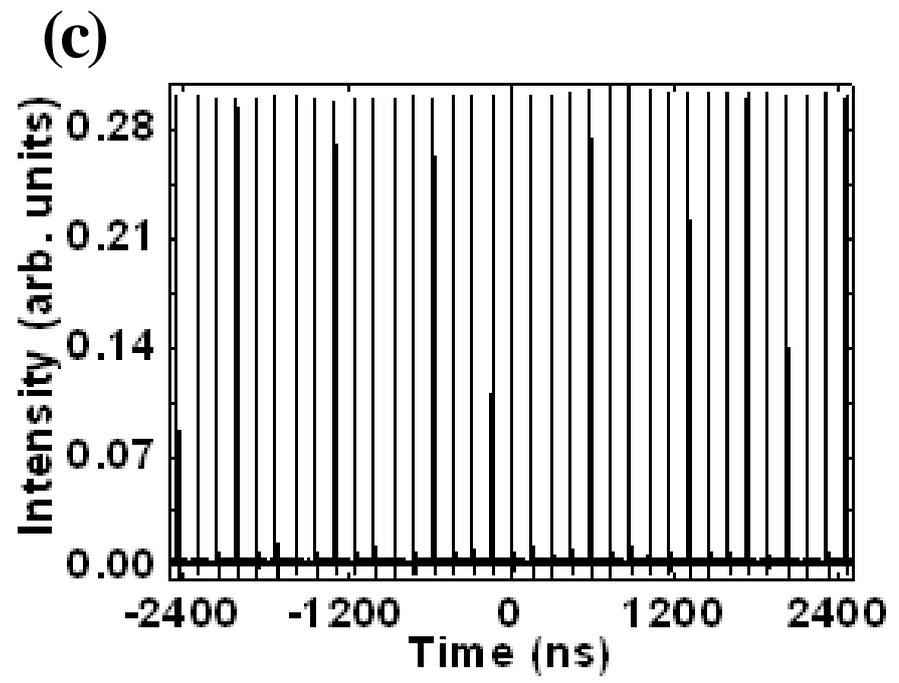

Fig.3 (c): Han Zhang et. al.

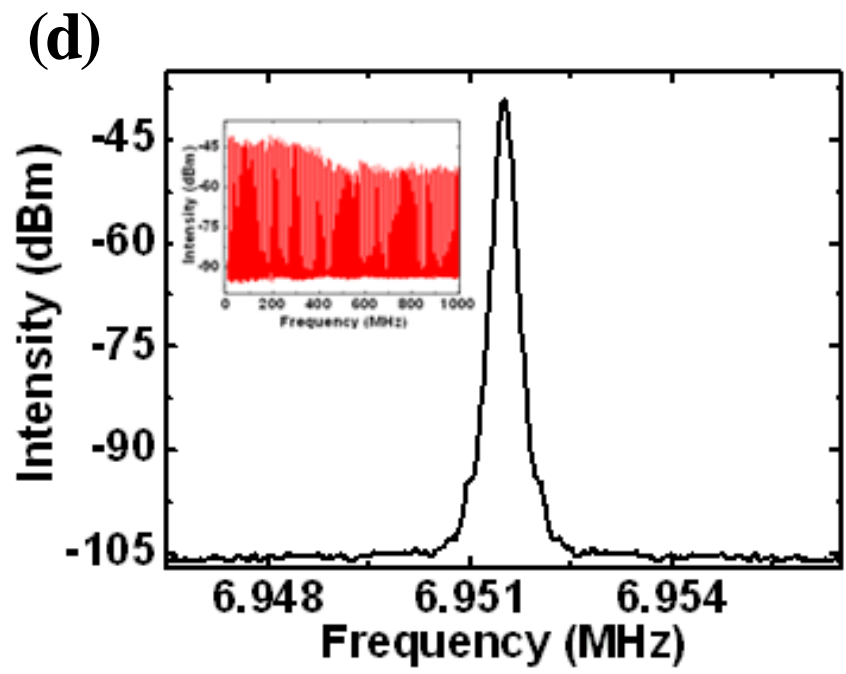

Fig.3 (d): Han Zhang et. al.